\documentstyle[12pt,psfig]{article}
\def\Lc{\Lambda_c}
\def\Lb{\Lambda_b}
\def\L{\Lambda}
\def\als{\mbox{$\alpha_s$}}
\def\ARL{\mbox{A}_{{ RL}}}

\def\mz{m_{ Z}}
\def\mq{m_{ Q}}
\def\md{m_{ D}}
\def\MA{{\cal M}}
\def\M0{{\cal M}_o}
\def\sl{s_{\Lambda_b}}

\begin{document}


\begin{center}
{\large\bf $\Lb$ POLARIZATION IN THE  $Z$ BOSON DECAYS.}
\vspace{10mm}

V.A.~Saleev
\vspace{2mm}

Samara State University, Samara, Russia
\vspace{1cm}
\end{center}

\begin{abstract}
In the framework of the perturbative QCD and the diquark model
of baryons we have obtained the fragmentation functions for heavy
quark that split into polarized $\Lb$ baryons. We predict the
longitudinal polarization asymmetry for the prompt $\Lb$ produced
in $e^+e^-$ annihilation at the $Z$ resonance and estimate
the spin-1/2 and spin-3/2 beauty baryon production rate.
\end{abstract}

\section*{Introduction}
The heavy baryon production in $e^+e^-$ annihilation is an
increasingly important subject to study. The recent
measurements near the threshold of the heavy quark production,
as well as at the $Z$ resonance, give us the information
about heavy baryon masses, life times,
decay modes \cite{1,2,3} and polarizations \cite{4,5}.

The $b$ quarks produced in $e^+e^-$ annihilation at the $Z$ peak are
strongly polarized. Accordingly to the Standard Model the right-left
longitudinal asymmetry is $\ARL^b=-0.94$. This polarization is
expected \cite{6} to be only slightly (2\%) reduced by the hard gluon
emission
before the hadronization into baryons. Therefore the
measurement of the $\Lb$ polarization is very important for study the
details of the hadronization processes. The first measurement \cite{4}
gave an intriguingly small value $\ARL^{\Lb}=-0.23^{+0.26}_{-0.23}$,
which contradicts the prediction that the large part of the
initial polarization of the quark to transfer to the $\Lb$ baryon.

The data on the $\Lb$ polarization may be described in part
($\ARL^{\Lb}\approx -0.68$) by means
of a simple depolarization model, based on the cascade $\Lb$ partial
production in the decays of higher spin beauty baryon
produced in the $Z$ decays \cite{7}.

In this paper we propose the depolarizing mechanism occurring during
the prompt $\Lb$ production via $b$ quark fragmentation. Our approach
 base on the perturbative QCD, the quark-diquark model of the
baryons \cite{8} and the nonrelativistic approximation, which is used
successfully for the description of the large distance effects in the
heavy quarkonium production \cite{9}. The same approach have been
discussed in the our previous paper \cite{10}, where the predictions
for the  doubly heavy   baryon production via the heavy quark
fragmentation have been presented.

\section{The prompt $\Lb$ production in $Z$ decays}
The fragmentation function $D_{b\to\Lb}(z,\mu)$ at the initial scale
$\mu=\mu_o=\mq+2\md$ for the production of the $\Lb$ baryon,
containing the $Q=b$ quark and the scalar diquark $D=(ud)$, is given
by the following expression \cite{10,11}:
\begin{equation}
D_{b\to \Lb}(z,\mu_o)=\frac{1}{16\pi^2}\int_{s_{min}}^{\infty}
ds \lim_{q_o\to\infty}\frac{|\MA|^2}{|\M0|^2},
\end{equation}
where $\MA$ is the matrix element for the production $\Lb$ baryon and
antidiquark $\bar D$ with the total four-momentum $q=p+q'$ and invariant
mass $s=q^2$ in the $Z$ decay ($Z\to\Lb\bar D\bar b$), $\M0$ is the
matrix element for the production of a real $b$ quark with the same
three-momentum $\vec q$. The lower limit in the integral (1) is
$$ s_{min}=\frac{M^2}{z}+\frac{\md^2}{1-z},$$
where $M=\mq+\md$ is the baryon mass, $\mq$ is the heavy quark mass,
$\md$ is the diquark mass.

The amplitude $\M0$ may be presented as follows:
\begin{equation}
\M0(Z\to b\bar b)=\bar U(q)\Gamma^{\alpha} V(\bar q)
\varepsilon_{\alpha}(Z),
\end{equation}
where $\varepsilon_{\alpha}(Z)$ is the polarization four-vector
of the $Z$ boson and
$$\Gamma^{\alpha}=-\frac{ig}{2\cos{\theta_{ W}}}\gamma^{\alpha}
\left (\mbox{C}_{ V}^b-\mbox{C}_{ A}^b\gamma^5\right )$$
is the quark-boson vertex.

In the axial gauge for the gluon propagator
associated with the four-vector $n=(1,0,0,-1)$:
$$d_{\alpha\beta}(k)=-g_{\alpha\beta}+
\frac{k_{\alpha}n_{\beta}+k_{\beta}n_{\alpha}}{(kn)},$$
the fragmentation contribution comes only from the Feynman diagram
shown in Fig.1. After some obvious simplification we have obtained the
matrix element of the $Z$ decay into $\Lb$ baryon \cite{10}:
\begin{eqnarray}
&&\MA (Z\to\Lb\bar b\bar D)=\frac{\Psi(0)}{\sqrt{2\md}}g_s^2
\frac{4\delta^{ij}}{3\sqrt{3}}\frac{F_s(k^2)}{(s-\mq^2)^2}\cdot\nonumber\\
&&
2\bar U(p)\left [-M(\hat q+\mq)+(s-\mq^2)\left (\frac{np}{nk}\right )
\right ]\Gamma^{\alpha}V(\bar q)\varepsilon_{\alpha}(Z)
\end{eqnarray}
where $g_s^2=4\pi\alpha_s$, $4\delta^{ij}/3\sqrt{3}$ is the color
factor of the diagram,
$F_s(k^2)$ is the form factor of the vertex $g^*\to D\bar D$,
 $\Psi(0)$ is the $\Lb=(QD)$ wave function at the origin in the
quark-diquark approximation. The gluon coupling to scalar diquarks was
used in the following form \cite{8}:
\begin{equation}
S_{\nu}^a=-ig_sT^a(q'-p_D)_{\nu}F_s(k^2),
\end{equation}
where $T^a=\lambda^a/2$ are Gell-Mann matrices, $k=q'+p_D$ is the
gluon four-momentum, the spinor $\bar U(p)$ describes $\Lb$ baryon,
$p=p_Q+p_D$, $p_D=rp$, $p_Q=(1-r)p$, $r=\md/M$.
The scalar diquark form factor at $k^2>0$ may be
parameterized as in ref.\cite{12}, where the authors fit successfully the
angular distributions of the baryons in the processes
$\gamma\gamma\to p\bar p,\L\bar \L$ and the widths of $J/\psi\to p\bar
p,\L\bar \L$:
\begin{equation}
F_s(k^2)=\frac{Q_s^2}{Q_s^2-k^2},
\end{equation}
where $Q_s^2=3.22$ GeV$^2$ and form factor is restricted to value
smaller than 1.3. We use also the parameterization (5) with the fixed
full width at half maximum $\Gamma\approx 0.8\mbox{ GeV}^2$.
In the
both cases there are no singularities in the physical region of the
virtuality of the fragmenting quark or the square of the gluon
four-momentum.

The phenomenological diquark form factor parameterization (5)
corresponds to the elastic vertex $g^*\to D\bar D$ \cite{12}.
However, in the fragmentation processes the contribution from the
inelastic vertex $g^*\to D\bar u\bar d$ may be dominant. There are no
any information about inelastic diquark form factors. That is why we
will use parameterization (5) to describe spin effects in the $\Lb$
production. As it will be shown, the choice of the form factor
parameterization is important only for the prediction the absolute
values of the heavy baryon production rates, but the spin asymmetry
doesn't depend on a diquark form factor in the kinematic region, which
is studied.

The reduced mass of the heavy quark -- diquark system
 is the same order as one for the
system of the two charmed quarks and we can hope that the calculation
of the parameter $\Psi(0)$ using the potential approach, will be
well-grounded and may be done with the quark-diquark interaction
potential, which has been fixed previously in calculating the heavy
baryon mass spectrum \cite{13}.

The energy distribution of the right (R) or the left (L) longitudinally
polarized $\Lb$ baryons produced in the $Z$ decays reduces
at leading order in $\als$ to
\begin{equation}
\frac{d\Gamma^{R,L}}{dz}(Z\to \Lb X)=\Gamma(Z^o\to b\bar b)
D_{b\to \Lb}^{R,L}(z,\mu),
\end{equation}
where $$z=\frac{p_0+p_3}{q_0+q_3},$$
$p=(p_0,0,0,p_3)$ is 4-momentum of the $\Lb$ ,
$q=(q_0,0,0,q_3)$ is 4-momentum of the fragmenting b-quark
and $\mu\approx \mz/2$. Let us define
\begin{equation}
\Delta D_{b\to\Lb}(z,\mu)=D_{b\to\Lb}^R(z,\mu)-D_{b\to\Lb}^L(z,\mu).
\end{equation}
The unpolarized fragmentation function is
\begin{equation}
 D_{b\to\Lb}(z,\mu)=D_{b\to\Lb}^R(z,\mu)+D_{b\to\Lb}^L(z,\mu).
\end{equation}
The longitudinal asymmetry $\ARL^{\Lb}$ may be presented as follows:
\begin{equation}
\ARL^{\Lb}(z,\mu)=\frac{\Delta D_{b\to\Lb}(z,\mu)}{D_{b\to\Lb}(z,\mu)}.
\end{equation}
In the limit of $\mz^2>>s,M^2$ we have got
\begin{equation}
\mbox{Tr}\left [\hat q\Gamma_{\alpha}\hat{\bar q}\Gamma_{\beta}\right ]
\left (-g^{\alpha\beta}+{Z^{\alpha}Z^{\beta}}/{\mz^2}\right )
=\mz^2(\mbox{C}_{ V}^{b2}+\mbox{C}_{ A}^{b2})
\left [\frac{g^2}{4\cos ^2{\theta_{ W}}}\right ],
\end{equation}
\begin{equation}
\mbox{Tr}\left [\hat q\gamma_5\Gamma_{\alpha}\hat{\bar q}\Gamma_{\beta}\right ]
\left (-g^{\alpha\beta}+{Z^{\alpha}Z^{\beta}}/{\mz^2}\right )
=-2\mz^2\mbox{C}_{ V}^b\mbox{C}_{ A}^b
\left [\frac{g^2}{4\cos ^2{\theta_{ W}}}\right ].
\end{equation}
Taking into account that
the four-vector of the $\Lb$ polarization in
the states with the different longitudinal polarization is
$\sl^{R,L}=\pm (p_3/M,0,0,p_0/M)$ and using the following exact formulae
$$2pk=s-\mq^2,\quad k^2=r(s-\mq^2),$$
$$ 2pq=s-\mq^2+2M^2(1-r),\quad
2Zq=\mz^2+s-\mq^2,$$
we can reduce the squared matrix elements to the form, which contains
expressions (10) or (11).

Omitting the next details, we write here the obtained results for
the unpolarized fragmentation function at the scale $\mu=\mu_o$:
\begin{eqnarray}
D_{b\to\Lb}(z,\mu_o)&=&D_o\int_{s_{ min}}^{\infty}\frac{ds}{M^2}
F_s^2(k^2)\left (\frac{M^2}{s-\mq^2}\right )^4\nonumber\\
&&\left [\mbox{d}_o+\mbox{d}_1\left (\frac{s-\mq^2}{M^2}\right )
+\mbox{d}_2\left (\frac{s-\mq^2}{M^2}\right )^2\right ],
\end{eqnarray}
where
\begin{equation}
D_o=\frac{32\als^2|\Psi(0)|^2}{27rM^3},
\end{equation}
\begin{equation}
\mbox{d}_o=4(1-r),\quad
\mbox{d}_1=\frac{1-(4-r)z-(1-r)z^2}{1-z(1-r)},\quad
\mbox{d}_2=\frac{z^3}{(1-z(1-r))^2},
\end{equation}
The result for the $\Delta D_{b\to\Lb}(z,\mu_o)$ may be written as for
the $D_{b\to\Lb}(z,\mu_o)$ but we don't present it here because of
it's unwieldy.

The recalculating of the fragmentation functions $D^{RL}_{\Lb}(z,\mu_o)$
from the start point $\mu=\mu_o$ to the $\mu>\mu_o$ may be done using
Gribov-Lipatov-Altareli-Parisi (GLAP) evolution equations
\begin{equation}
\mu\frac{\partial D}{\partial\mu}(z,\mu)=\int_{z}^1\frac{dy}{y}
P_{Q\to Q}\left (\frac{z}{y},\mu\right )D(y,\mu),
\end{equation}
where $P_{Q\to Q}$ is the splitting function at leading order in $\als$:
\begin{equation}
P_{Q\to Q}(x,\mu)=\frac{4\als (\mu)}{3\pi}\left
(\frac{1+x^2}{1-x}\right )_+,
\end{equation}
$$f(x)_+=f(x)-\delta(1-x)\int_0^1f(x')dx'.$$
Note that at leading order in $\als$ one has
$$\int P_{Q\to Q}(z,\mu)dz=0,$$
and the evolution equation implies that the fragmentation probability
$$P_{b\to\Lb}=\int D_{b\to\Lb}(z,\mu)dz$$
as well as the total asymmetry
$\ARL^{\Lb}$ don't evolve with the scale $\mu$. However,  the
$z-$dependence of the $\ARL^{\Lb}(z,\mu)$ changes drastically when the
scale $\mu$ growth from $\mu=\mu_o$ to $\mu\approx\mz/2$.

First of all, we discuss here the results concerning spin effects in the
prompt $\Lb$ production, which don't depend on  the value of $\Psi(0)$.
Figs. 2-4 demonstrate the dependence of the polarization asymmetry
$\ARL^{\Lb}$ on the diquark mass, the form factor parameterization, $z$
and QCD evolution scale $\mu$.

We can see in Fig.2 that $\ARL^{\Lb}\ne\ARL^b$ at the realistic values
of the diquark mass ($\md=0.6-0.9$ GeV corresponds to
$|\ARL^{\Lb}|=0.90-0.87$)  and only at the limit of $\md\to 0$ one has
$\ARL^{\Lb}\approx\ARL^b=-0.94$. Our result corrects the assumption,
which is used usually, that the $\Lb$ baryon retain a large part of
the initial polarization of the b quark \cite{14}.
The predictions for the asymmetry as a function of $z$ strongly
depends on the diquark mass at $\mu=\mu_o$, but at the large $\mu$
this dependence vanishes (Fig.3).

$\ARL^{\Lb}(z,\mu)$ may be sensitive to the choice of the form factor
parameterization at the small $\mu$ (see Fig.4), however at the scale
$\mu=m_Z/2$ the results corresponding the different form factor
parameterizations are equal. The measurement of the
$\ARL^{\Lb}$ versus $z$ at the different $\mu$ is a exact test of the
hadronization model. However, it may be done experimentally if we can to
separate the prompt $\Lb$ production in the b quark fragmentation from
the $\Lb$ produced in the cascades: $b\to\Sigma_b^*,...$ plus
the hadronical decays $\Sigma_b^*,...\to \Lb$. This opportunity was
shown recently in the study of the heavy quarkonium production at FNAL
\cite{15}, where the vertex detector technique was used.

The Figs.5  show our results for the $\ARL^{\Lc}$
versus $z$ at the different $\mu=\mu_o,\mz/2$ and $m_D=0.6$ GeV.
We have obtained
$\ARL^{\Lc}=-0.58$ opposite to $\ARL^{c}=-0.67$.
The probability of the heavy quark spin turning during the hadronization is
about 13\% for the c-quark and only 4\% for the b-quark. The both
values are
independent from the diquark form factor and were obtained at
$m_D=0.6$ GeV. We see that the simple rule ($\sim m_D/m_Q$) is
satisfied.

\section{The cascade $\Lb$ production in the $Z$ decays}
Using the value of the radial part of the $\Lb$ baryon wave function
 $|R(0)|^2=0.8$ GeV$^3 \quad (\alpha_s=0.4, \quad m_D=0.6 \mbox{ GeV})$,
which was calculated
in the quark-diquark approximation \cite{13}, we have found the total
probability of the b quark fragmentation into the $\Lb$:
$P_{b\to\Lb}\approx 2.5\cdot 10^{-3}$ in the case of the form
factor (5) and
$P_{b\to\Lb}\approx 2.5\cdot 10^{-2}$ if the singularities are
restricted using the fixed width $\Gamma$.  In spite of this deference
the shapes of the fragmentation functions $D_{b\to\Lb}(z,\mu)$ are the
same for the both form factors (Fig. 6).
The obtained value $P_{b\to\Lb}$
gives the number of $\Lb$ per the $Z$ decay which is  smaller than
it was measured at LEP \cite{3}. We conclude that the main part of
the $\Lb$ baryons are produced in the cascade processes via the decays
of the higher spin $b$ baryons. The DELPHI collaboration has presented
evidence for $\Sigma^*_b$ baryon production \cite{16} and measured the
rate to be $P_{b\to\Sigma_b^*}=4.8\pm 1.6\% $,
the total $b$ baryon production rate to be $P_{b\to baryon}=11.5\pm 4.0\%$
\cite{2}.

We can estimate the fragmentation probability for the $b$ quark that
split into $\Sigma_b^*$ baryon in the our approach. However, as it was
demonstrated in the case of the $\Lb$ baryon production, the predicted
absolute production rate is strongly depends on diquark form factor.
It will be more reliable to calculate
the ratio of the fragmentation probabilities for the $b$ quark
that split into
vector diquark states to scalar diquark states, using deferent form
factors. As it was shown in ref. \cite{7} this parameter (called $A$)
controls the $\Lb$ polarization in the cascade production.

The gluon coupling to axial vector diquarks is presented by the
following expression:
$$V_{\mu}^b =ig_sT^b \biggl \{ \varepsilon^*_D\varepsilon_{\bar D}
(q'-p_D)_{\mu}F_1(k^2)-
 [(q'\varepsilon_{\bar D})\varepsilon^*_{D\mu}-
(p_D\varepsilon^*_D)\varepsilon_{\bar D\mu}] F_2(k^2)-$$
\begin{equation}
(\varepsilon^*_D q')(\varepsilon_{\bar D} p_D)
(q'-p_D)_{\mu}F_3(k^2)\biggl \},
\end{equation}
where   $\varepsilon^*_D, \varepsilon_{\bar D}$ are the diquark
polarization four-vectors.
$F_1, F_2$ and $F_3$ are form factors depending on the momentum
transfer squared $k^2=(q'+p_D)^2$.
The same as in ref. \cite{12} we put for $k^2>0$:
\begin{equation}
F_1(k^2)=\Biggl({Q_V^2\over{Q_V^2-k^2}}\Biggr)^2,
\end{equation}
$$F_2(k^2)=(1+\kappa)F_1(k^2),\qquad F_3(k^2)=0$$
where $Q_V=1.50$ GeV$^2$ and $\kappa$ being the anomalous chromomagnetic
moment of the vector diquark. The anomalous magnetic moment of the
(ud) vector diquark was
determined in ref.\cite{12}, but it is not obviously that the
chromomagnetic and magnetic moments are equal. We will choose
$\kappa=0$ for the simplicity.

The amplitude for heavy quark fragmentation into spin-3/2 baryons,
corresponding to fusion of the heavy quark  and the axial vector diquark,
can be written as a sum of two parts, which are proportional to $F_1$
and $F_2$ form factors:
\begin{equation}
{\MA}_{3/2}={\MA}_{3/2}^1+{\MA}_{3/2}^2,
\end{equation}

\begin{equation}
{\MA}_{3/2}^1=\frac{\Psi(0)}{\sqrt{2m_D}}
g_s^2{4\delta^{ij}\over{3\sqrt{3}}}{F_1(k^2)\over{(s-m_Q^2)^2}}
\end{equation}
$$
2\bar\Psi_{\mu}(p_Q)\varepsilon_{\bar D\mu}
\Biggl [-M(\hat q+m_Q)+(s-m_Q^2){(np)\over{(nk)}}\Biggr ]
\Gamma^{\alpha}V(\bar q)\varepsilon_{\alpha}(Z),
$$
\begin{equation}
{\MA}_{3/2}^2=\frac{\Psi(0)}{\sqrt{2m_D}}
g_s^2{4\delta^{ij}\over{3\sqrt{3}}}{F_2(k^2)\over{(s-m_Q^2)^2}}
\end{equation}
$$
{1\over r}\bar\Psi_{\sigma}(p_Q)\varepsilon_{\bar D\lambda}
\Biggl [q^{\sigma}\gamma^{\lambda}(\hat q+\mq)+
\frac{s-\mq^2}{(kn)}(k^{\lambda}n^{\sigma}-k^{\sigma}n^{\lambda})
\Biggr ]\Gamma^{\alpha} V(\bar q)\varepsilon_{\alpha}(Z).$$
Here, the $\bar\Psi_{\mu}(p)$ is the Rarita-Schwinger spinor for the
spin-3/2 baryons. The procedure of the calculation for the fragmentation
function $D_{b\to\Sigma_b^*}(z,\mu)$ is the same as for the $\Lb$
baryons.

Using the set of parameters:
 $|R(0)|^2=0.8 \mbox{GeV}^3, \quad \alpha_s=0.4, \quad m_D=0.6 \mbox{ GeV}$,
$\kappa=0$, we have found that
the ratio  $A$ of the fragmentation probabilities for the $b$ quark
that split into
vector diquark states to scalar diquark states is approximately equal
to $5\pm 1$ independently on the choice of the diquark form factor.
Accordingly the formula (5.10) ref.\cite{7} one has
\begin{equation}
\ARL^{\Lb}=\frac{1+(1+4 w_1)A/9}{1+A} \ARL^b,
\end{equation}
where the parameter $w_1$ gives the probability that spin-1 diquark
has maximum angular momentum $j^3=\pm 1$ along the fragmentation axis.
The authors ref. \cite{7} used $A=0.45, \quad w_1=0$, as it is
motivated by the Lund fragmentation model \cite{16b}, and obtained
that $\ARL^{\Lb}=-0.68$. If we put $A=5$ in (22), as it follows from our
results, we find that $\ARL^{\Lb}=-0.24$ which is in a good agreement
with the value of the measurement.

In conclusion we want note that the prediction of the cascade
model \cite{7} for the $\ARL^{\Lb}$  in the $Z$ decays
was obtained using the assumption that the turning of the $b$ quark
spin may be ignored, as is appropriate to the heavy quark
approximation. Our results show that the
probability of the $b$ quark overturning during the hadronization
may equal to 4-7 \% and the more careful analysis of the depolarization
in the cascade $\Lb$ production is needed.

When our work was completed the result of the computing  the
heavy baryon production rates at CLEO and LEP  energies, which was obtained
a similar way, has been presented \cite{17}.

\section*{Acknowledgments}
We are grateful to  S.~Gerasimov and A.~Likhoded
for discussions the problems of the heavy
baryon physics and G.~Goldstein for the valuable information on
obtained results.

\section*{Figure captions}
\begin{enumerate}
\item Diagram used for description of the process
      $Z^o\to\Lb\bar b\bar D$.
\item The asymmetry $\ARL^{\Lb}$ as a function of the ratio
      $\md/M$.
\item The asymmetry $\ARL^{\Lb}$ as a function of $z$
      at $\mu=\mu_o$ (curves 1 and 2) and $\mu=\mz/2$ (curves 3 and 4).
      The solid lines correspond to $\md=0.9$ GeV, the dotted lines
      correspond to $\md=0.6$ GeV.
\item
The asymmetry $\ARL^{\Lb}$ as a function of  $z$
      at $\md=0.6$ GeV and $\mu=\mu_o$ (curves 1 and 2), $\mz/2$
(curves 3 and 4).
      The solid lines correspond to diquark form factor
parameterization (5),
      the dotted lines correspond to formula (5) with the fixed width
      $\Gamma=0.8$ GeV$^2$.
\item
The asymmetry $\ARL^{\Lc}$ as a function of $z$
      at $\md=0.6$ Ē' and $\mu=\mu_o$ (curve 1), $\mz/2$ (curve 2).
\item
The fragmentation function $D_{b\to\Lb}(z,\mu)$ normalized to
unity at $\mu=\mu_0$ (curves 1 and 2) and $\mu=\mz/2$ (curves 3 and 4).
The solid and dotted lines correspond
to the different form factor parameterizations.
\end{enumerate}

\begin{figure}[p]
\psfig{figure=pic1.ps,height=10cm,width=10cm,%
        bbllx=2cm,bblly=10cm,bburx=12cm,bbury=20cm}%
\vspace*{1.0cm}
\caption
{Diagram used for description of the process
      $Z^o\to\Lb\bar b\bar D$.}
\end{figure}
\begin{figure}[p]
\psfig{figure=pic2.ps,height=10cm,width=10cm,%
        bbllx=2cm,bblly=10cm,bburx=12cm,bbury=20cm}%
\vspace*{1.0cm}
\caption
{The asymmetry $\ARL^{\Lb}$ as a function of the ratio
 $\md/M$.}
\end{figure}
\begin{figure}[p]
\psfig{figure=pic3.ps,height=10cm,width=10cm,%
        bbllx=2cm,bblly=10cm,bburx=12cm,bbury=20cm}%
\vspace*{1.0cm}
\caption
{The asymmetry $\ARL^{\Lb}$ as a function of $z$
 at $\mu=\mu_o$ (curves 1 and 2) and $\mu=\mz/2$ (curves 3 and 4).
TThe solid lines correspond to $\md=0.9$ GeV, the dotted lines
 correspond to $\md=0.6$ GeV.}
\end{figure}
\begin{figure}[p]
\psfig{figure=pic4.ps,height=10cm,width=10cm,%
        bbllx=2cm,bblly=10cm,bburx=12cm,bbury=20cm}%
\vspace*{1.0cm}
\caption
{
The asymmetry $\ARL^{\Lb}$ as a function of  $z$
      at $\md=0.6$ GeV and $\mu=\mu_o$ (curves 1 and 2), $\mz/2$
(curves 3 and 4).
      The solid lines correspond to diquark form factor
parameterization (5),
      the dotted lines correspond to formula (5) with the fixed width
      $\Gamma=0.8$ GeV$^2$.}
\end{figure}
\begin{figure}[p]
\psfig{figure=pic5.ps,height=10cm,width=10cm,%
        bbllx=2cm,bblly=10cm,bburx=12cm,bbury=20cm}%
\vspace*{1.0cm}
\caption
{The asymmetry $\ARL^{\Lc}$ as a function of $z$
      at $\md=0.6$ Ē' and $\mu=\mu_o$ (curve 1), $\mz/2$ (curve 2).}

\end{figure}
\begin{figure}[p]
\psfig{figure=pic6.ps,height=10cm,width=10cm,%
        bbllx=2cm,bblly=10cm,bburx=12cm,bbury=20cm}%
\vspace*{1.0cm}
\caption
{
The fragmentation function $D_{b\to\Lb}(z,\mu)$ normalized to
unity at $\mu=\mu_0$ (curves 1 and 2) and $\mu=\mz/2$ (curves 3 and 4).
The solid and dotted lines correspond
to the different form factor parameterizations.}
\end{figure}

\end{document}